\documentclass[12pt]{article}
\input{epsf}
\newcommand{\kart}[4]{\begin{figure}[#1]\begin{center}\leavevmode%
\epsfysize=#2\epsffile{#3.ps}\end{center}\caption{#4}\end{figure}}

\author{S. V. Krasnikov\thanks{Email: \it redish@pulkovo.spb.su}\\}
\title{Toward a Traversable Wormhole}
\begin{document}
\maketitle
\begin{abstract}
In this talk I discuss pertinence of the wormholes to the problem of
circumventing the light speed barrier and present a specific class of
wormholes. The wormholes of this class are static and have
arbitrarily wide throats, which makes them traversable. The matter
necessary for these spacetimes to be solutions of the Einstein
equations is shown to consist of two components, one of which
satisfies the Weak energy condition and the other is produced by
vacuum fluctuations of neutrino, electromagnetic (in dimensional
regularization), and/or massless scalar (conformally coupled) fields.
\end{abstract}

\section*{Wormholes and their application to hyper-fast travel}

Wormholes are geometrical structures connecting two more or less flat
regions of a spacetime. This of course is not a rigorous definition,
but, strange though it may seem, there is no commonly accepted
rigorous definition of the wormhole yet. Normally, however, by a
wormhole a spacetime is understood resembling that obtained by the
following manipulation: 
\begin{enumerate}
\item Two open balls are removed each from a piece of approximately
flat 3-space (the vicinities of thus obtained holes 
we shall call \emph{mouths} of the wormhole); 
\item The boundaries (2-spheres) of the holes are glued together, and
the junction is smoothed. In the process of 
smoothing a kind of tube arises interpolating the spheres. We shall
call this tube the \emph{tunnel} and its narrowest part 
the \emph{throat}.
\end{enumerate}
The resulting object (its two-dimensional version to be precise) is
depicted in Fig.~1. If in the course of evolution the 
spacetime surrounding such an object remains approximately flat
(which may not be the case, since flatness of each 3-dimensional
section does not guarantee that the 4-dimensional space formed by
them is also flat) we shall call the 
object a wormhole.
Wormholes arise in a natural way in general relativity. Even one of
the oldest and best-studied solutions of the Einstein 
equations --- the Schwarzschild spacetime --- contains a wormhole,
which was found at least 80 years ago (Flamm, 
1916). This wormhole (also known as the Einstein-Rosen bridge)
connects two asymptotically flat regions (`two 
universes'), but being non-static is useless in getting from one of
them to the other (see below). 

Depending on 
how the vicinities of the mouths are extended to the full spacetime
the wormholes fall into two categories (Visser, 1995): 
It may happen that the mouths cannot be connected by any curve except
those going through the tunnel (as it takes 
place in the Einstein-Rosen bridge). Such wormholes are called
\emph{inter-universe}.
 A simplest static spherically symmetric
inter-universe 
wormhole can be described (Morris, 1988) by a manifold R$^2\times$
S$^2$ endowed with the metric 
\begin{equation} 
ds^2 = - e^{2\Phi} dt^2 + dr^2/ (1- b/r) + r^2 (d\theta^2 +
\sin^2\theta d\phi^2), 
\end{equation}
where $r\in(-\infty,\infty)$ (note this possibility of negative $r$ ,
it is the characteristic feature of the wormholes), $\Phi(r)\to 0$
and 
$b(r)/r\to 0$, when $r \to\pm\infty$.
Alternatively as shown in Fig.~1 it may happen that there are curves
from one mouth to another lying outside the 
wormhole. 
\kart{h}{2in}{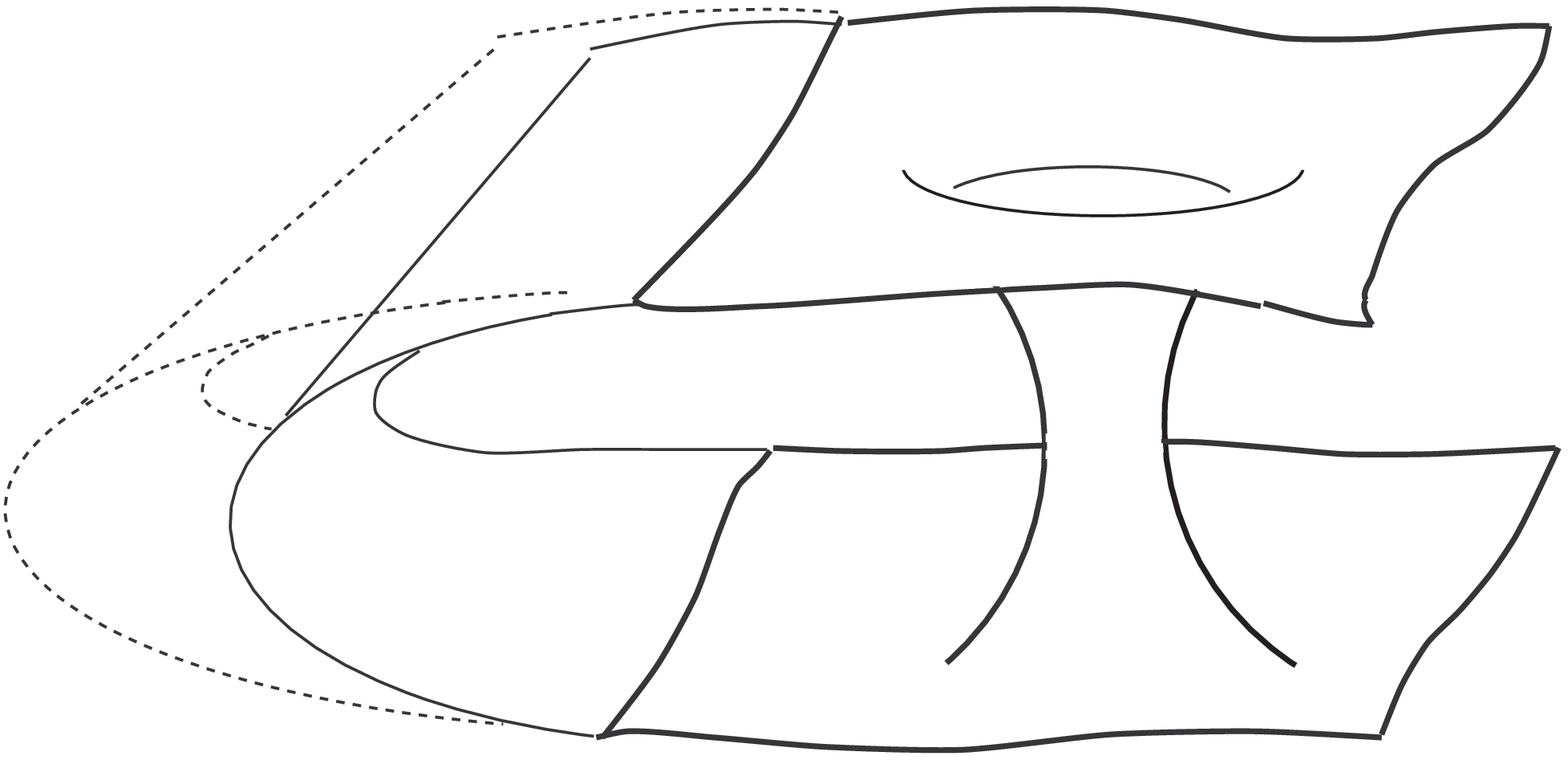}{The sketch of a wormhole with the mouths in
motion. One dimension (corresponding to the coordinate $\theta$) is 
omitted. The ways in which the upper and the lower parts are glued at
$t = 0$ and at $t = 1$ are depicted by thin solid lines and by 
dashed lines respectively. Though the geometry of the wormhole does
not change, the distance (as measured in the outer, flat 
space) between mouths increases with time.}
Such a wormhole connects distant parts of a `single'
universe and is called \emph{intra-universe}. Though intra-universe
wormholes are in a sense more interesting most papers deal with
inter-universe ones, since they are simpler. It 
does not matter much, however. The distant regions of the `universes'
are taken to be approximately flat. And it is 
usually implied that given an inter-universe wormhole we can as well
build an intra-universe one by simply gluing these 
distant regions in an appropriate way.

It is stable intra-universe wormholes that are often used for
interstellar travel in science fiction (even though they are 
sometimes called `black holes' there). Science fiction (especially
Sagan's novel \emph{Contact}) apparently acted back on 
science and in 1988 Morris and Thorne pioneered investigations
(Morris, 1988) of what they called \emph{traversable 
wormholes} --- wormholes that can be (at least in principle)
traversed by a human being. It is essential in what follows 
that to be traversable a wormhole should satisfy at least the
following conditions: 
\begin{itemize}
\item[(C1).] It should be sufficiently stable. For example the
Einstein-Rosen bridge connects two asymptotically flat 
regions (and so it is a wormhole), but it is not traversable --- the
throat collapses so fast that nothing (at least 
nothing moving with $v\leq c$) can pass through it.
\item[(C2).] It should be \emph{macroscopic}. Wormholes are often
discussed [see (Hoch\-berg, 1997), for example] with the radius of 
the throat of order of the Plank length. Such a wormhole might be
observable (in particular, owing to its 
gravitational field), but it is not obvious (and it is a long way
from being obvious, since the analysis would 
inevitably involve quantum gravity) that any signal at all can be
transmitted through its tunnel. Anyway such a 
wormhole is impassable for a spaceship.
\end{itemize}
Should a traversable wormhole be found it could be utilized in
interstellar travel in the most obvious way. Suppose a 
traveler (say, Ellie from the above-mentioned novel) wants to fly
from the Earth to Vega. One could think that the trip 
(there and back) will take at least 52 years (by the terrestrial clocks) 
even if she 
moves at a nearly light speed. But if there is a wormhole 
connecting the vicinities of the Earth and Vega she can take a
short-cut by flying through it and thus make the round 
trip to Vega in (almost) no time. 
Note, however, that such a use of a wormhole would have had nothing
to do with circumventing the light barrier.  
Indeed, suppose that Ellie's start to Vega is appointed on a moment
$t = 0$. Our concern is with the time interval $\Delta t_E$ in 
which she will return to the Earth. Suppose that we know (from
astronomical observations, theoretical calculations, 
etc.) that if in $t = 0$ she (instead of flying herself) just emit a
photon from the Earth, this photon after reaching Vega 
(and, say, reflecting from it) will return back at best in a time
interval $\Delta t_p$. If we find a wormhole from the Earth to 
Vega, it would only mean that $\Delta t_p$ actually is small, or in
other words that Vega is actually far closer to the Earth than 
we think now. But what can be done if $\Delta t_p$ \emph{is} large
(one would hardly expect that traversable wormholes can be found 
for \emph{any} star we would like to fly to)? That is where the need
in hyper-fast transport comes from. In other words, the 
problem of circumventing the light barrier (in connection with
interstellar travel) lies in the question: how to reach a 
remote (i.~e.\ with the large $\Delta t_p$) star and to return back
sooner than a photon would have made it (i. e. in $\Delta t_E<\Delta
t_p$)? It 
makes sense to call a spaceship faster-than-light (or
\emph{hyper-fast}) if it solves this prolem. 
A possible way of creating hyperfast transport lies also in the use
of traversable wormholes (Krasnikov, 1998). Suppose 
that a traveler finds (or builds) a traversable wormhole with both
mouths located near the Earth and suppose that she 
can move the mouths (see Fig.1) at will without serious damage to the
geometry of the tunnel (which we take to be 
negligibly short). Then she can fly to Vega taking one of the mouths
with her. Moving (almost) at the speed of light she 
will reach Vega (almost) instantaneously by her clocks. In doing so
she rests with respect to the Earth insofar as the 
distance is measured through the wormhole. Therefore her clocks
remain synchronous with those on the Earth as far as 
this fact is checked by experiments confined to the wormhole. So, if
she return through the wormhole she will arrive 
back to the Earth almost immediately after she will have left it
(with $\Delta t_E\ll\Delta t_p$). 

\paragraph{Remark 1.} The above arguments are very close to those
showing that a wormhole can be transformed into a time 
machine (Morris, 1988), which is quite natural since the described
procedure is in fact the first stage of such 
transformation. For, suppose that we move the mouth back to the Earth
reducing thus the distance between the mouths 
(in the ambient space) by 26 light years. Accordingly $\Delta t_E$
would lessen by $\approx$26 yr and (being initially very small) 
would turn \emph{negative}. The wormhole thus would enable a
traveller to return before he have started. Fortunately, $\Delta t_E
\approx 0$ 
would fit us and we need not consider the complications (possible
quantum instability, paradoxes, etc.) connected with 
the emergence of thus appearing time machine.

\paragraph{Remark 2.} Actually two \emph{different} worlds were
involved in our consideration. The geometry of the world where only a
photon was emitted differs from that of the world where the wormhole
mouth was moved. A photon emitted in $ t = 0$ in 
the latter case would return in some $\Delta t_{p'} < \Delta t_E$ .
Thus what makes the wormhole-based transport hyper-fast is changing 
(in the causal way) the geometry of the world so that to make $\Delta
t_{p'} < \Delta t_E\ll\Delta t_{p}$.
 
Thus we have seen that a traversable wormhole can possibly be used as
a means of `superluminal' communication. True, 
a number of serious problems must be solved before. First of all,
where to get a wormhole? At the moment no good 
recipe is known how to make a new wormhole. So it is worthwhile to
look for `relic' wormholes born simultaneously 
with the Universe. Note that though we are not used to wormholes and
we do not meet them in our everyday life this 
does not mean by itself that they are an exotic rarity in nature (and
much less that they do not exist at all). At present 
there are no observational limits on their abundance [see
(Anchordoqui, 1999) though] and so it well may be that there 
are 10 (or, say, 10$^6$) times as many wormholes as stars. However,
so far we have not observed any. So, this issue 
remains open and all we can do for the present is to find out whether
or not wormholes are allowed by known physics. 

\section*{Can traversable wormholes exist?}

Evolution of the spacetime geometry (and in particular evolution of a
wormhole) in general relativity is determined via the Einstein
equations by properties of the matter filling the spacetime. This
circumstance may turn out to be fatal for wormholes if the
requirements imposed on the matter by conditions (C1,C2) are
unrealistic or conflicting. That the problem is grave became clear
from the very beginning: it was shown (Morris, 1988), see also
(Friedman, 1993), that under very general assumptions the matter
filling a wormhole must violate the Weak Energy Condition (WEC). The
WEC is the requirement that the energy density of the matter be
positive in any reference system. For a diagonal stress-energy tensor
$T_{ik}$ the WEC may be written as                              
\begin{equation}
{\rm WEC:}\qquad T_{00}\geq 0, \quad T_{00} + T_{ii}\geq 0, \qquad i = 1,
2, 3 
\end{equation}
Classical matter always satisfies the WEC (hence the name
\emph{`exotic'} for matter violating it). So, a wormhole can be 
traversable only if it is stabilized by some quantum effects.
Candidate effects are known, indeed [quantum effects can violate any
local energy condition (Epstein, 1965)].  
Moreover, owing to the non-trivial topology a wormhole is just a
place where one would expect WEC violations due to 
fluctuations of quantum fields (Khatsymovsky, 1997a). So, the idea
appeared (Sushkov, 1992) to seek a wormhole with 
such a geometry that the stress-energy tensor produced by vacuum
polarization is exactly the one necessary for 
maintaining the wormhole. An example of such a wormhole (it is a
Morris-Thorne spacetime filled with the scalar non-minimally coupled
field) was offered in (Hochberg, 1997). Unfortunately, the diameter
of the wormhole's throat was 
found to be of the Plank scale, that is the wormhole is non-traversable. 
The situation considered in (Hochberg, 1997) is of course very
special (a specific type of wormholes, a specific field, 
etc.). However arguments were cited [based on the analysis of another
energetic condition, the so called ANEC 
(Averaged Null Energy Condition)] suggesting that the same is true in
the general case as well (Flanagan, 1996, see 
also the literature cited there). So an impression has been formed
that conditions (C1) and (C2) are incompatible, and 
TWs are thus impossible.

\section*{Yes, it seems they can}

The question we are interested in is whether such macroscopic
wormholes exist that they can be maintained by the 
exotic matter produced by the quantum effects. To put it more
mathematically let us first separate out the contribution $T^Q_{ik}$
 of the `zero-point energy' to the total stress-energy tensor:
\begin{equation}
 T_{ik}  = T^Q_{ik} + T^C_{ik} . 
\end{equation}
In semiclassical gravity it is deemed that for a field in a quantum
state $|\Psi\rangle $ (in particular, $|\Psi\rangle$ may be a vacuum state)
$T^Q_{ik} = \langle\Psi| {\bf T}_{ik} |\Psi\rangle$, where ${\bf T}_{ik}$
 is the corresponding operator, and there are recipes for finding $T^Q_{ik}$
for given field, metric, 
and quantum state [see, for example, (Birrel, 1982)]. So, in formula
(3) $T^Q_{ik}$ and $T_{ik}$ are determined by the geometry of 
the wormhole and the question can be reformulated as follows: do such
macroscopic wormholes exist that the term $T^C_{ik}$ 
describes usual non-exotic matter, or in other words that $T^C_{ik}$
satisfies the Weak Energy Condition, which now can be 
written as
\begin{equation}
G_{00} - 8\pi T^Q_{00}\geq 0,\quad (G_{00} + G_{ii}) - 8\pi (T^Q_{00} +
T^Q_{ii} )\geq 0, \qquad i = 1, 2, 3.  
\end{equation}
(we used the formulas (2,3) here)?
One of the main problem in the search for the answer is that the
relevant mathematics is complicated and unwieldy. A 
possible way to obviate this impediment is to calculate $T^Q_{ik}$
numerically (Hochberg, 1997; Taylor 1997) using some 
approximation. However, the correctness of this approximation is in
doubt (Khatsymovsky, 1997b), so we shall not 
follow this path. Instead we shall study a wormhole with such a
metric that relevant expressions take the form simple 
enough to allow the analytical treatment.

The Morris-Thorne wormhole is not the unique static spherically
symmetric wormhole (contrary to what can often be 
met in the literature). Consider a spacetime R$^2\times$ S$^2$ with
the metric: 
\begin{equation}
ds^2 = \Omega^2 (\xi ) [- d\tau^2 +d\xi^2 + K^2 (\xi ) (d\theta^2 +
\sin^2\theta d\phi^2 ) ], 
\end{equation}
where $\Omega$ and $K$ are smooth positive even functions, $K= K_0
\cos\xi/L$ at $\xi \in (- L, L)$, $K_0 \equiv K (0)$ and $K$ is
constant at large 
$\xi.$ The spacetime is obviously spherically symmetric and static.
To see that it has to do with wormholes consider the 
case
\begin{equation}
\Omega \sim \Omega_0 \exp\{Bx\},\qquad              \mbox{\rm at large } \xi.
\end{equation}
The coordinate transformation
\begin{equation}
r \equiv B^{-1}\Omega_0 \exp{B\xi },  \quad         t \equiv Br\tau,        
\end{equation}
then brings the metric (5) in the region  $t < r$ into the form:
\begin{equation}
ds^2 = - dt^2 + 2t/r\,dt dr + [1- ( t/r)^2 ]dr^2 + (BK_0 r)^2
(d\theta^2 + \sin^2 \theta d\phi^2 ).  
\end{equation}
It is obvious from (7) that as $r$ grows the metric (5,8) becomes
increasingly flat (the gravitational forces corresponding 
to it fall as $1/ r$) in a layer $| t | < T$ ($T$ is an arbitrary
constant). This layer forms a neighborhood of the surface $\tau= t=
0$.  
But the spacetime is static (the metric does not depend on $\tau$).
So, the same is true for a vicinity of \emph{any} surface
$\tau=const$ 
. The spacetime can be foliated into such surfaces. So this property
(increasing flatness) holds in the whole spacetime, 
which means that it is a wormhole, indeed. Its length (the distance
between mouths as measured through the tunnel) is 
of order of $\Omega_0 L$ and the radius of its throat  $R_0 = \min (\Omega K)$.

The advantage of the metric (5) is that for the electro-magnetic,
neutrino, and 
massless conformally coupled scalar 
fields $T^Q_{ik}$ can be readily found (Page, 1982) in terms of
$\Omega, K$, and their derivatives [actually the expression contains 
also one unknown term (the value of $T^Q_{ik}$ for $\Omega = 1$), but
the more detailed analysis shows that for sufficiently large $\Omega$
this term can be neglected]. So, by using this expression,
calculating the Einstein tensor $G_{ik}$ for the metric (5) and 
substituting the results into the system (4) we can recast it [the
relevant calculations are too laborious to be cited here 
(the use of the software package \emph{GRtensorII} can lighten the
work significantly though)] into the form: 
\begin{equation}
 E_i \geq 0   \qquad        i =  0, 1, 2, 3, 
\end{equation}
where $E_i$ are some (quite complex, e.~g.~$E_0$ contains 40 terms;
fortunately they are not all equally important) 
expressions containing $\Omega$, $K$, and their derivatives and
depending on what field we consider. 
Thus if we restrict ourselves to wormholes (5), then to answer the
question formulated above all we need is to find out 
whether such $\Omega$ exist that it
\begin{enumerate}
\renewcommand{\theenumi}{{\roman{enumi}\/})}
\item has appropriate asymptotic behavior [see (6)],
\item  satisfies (9) for some field,  
\item delivers sufficiently large $R_0$.
\end{enumerate}
It turns out (Krasnikov, 1999) that for all three fields listed above
and for arbitrarily large $R_0$ such $\Omega$ do exist (an 
example is sketched in Fig.~2) and so the answer is positive.

\kart{h}{2in}{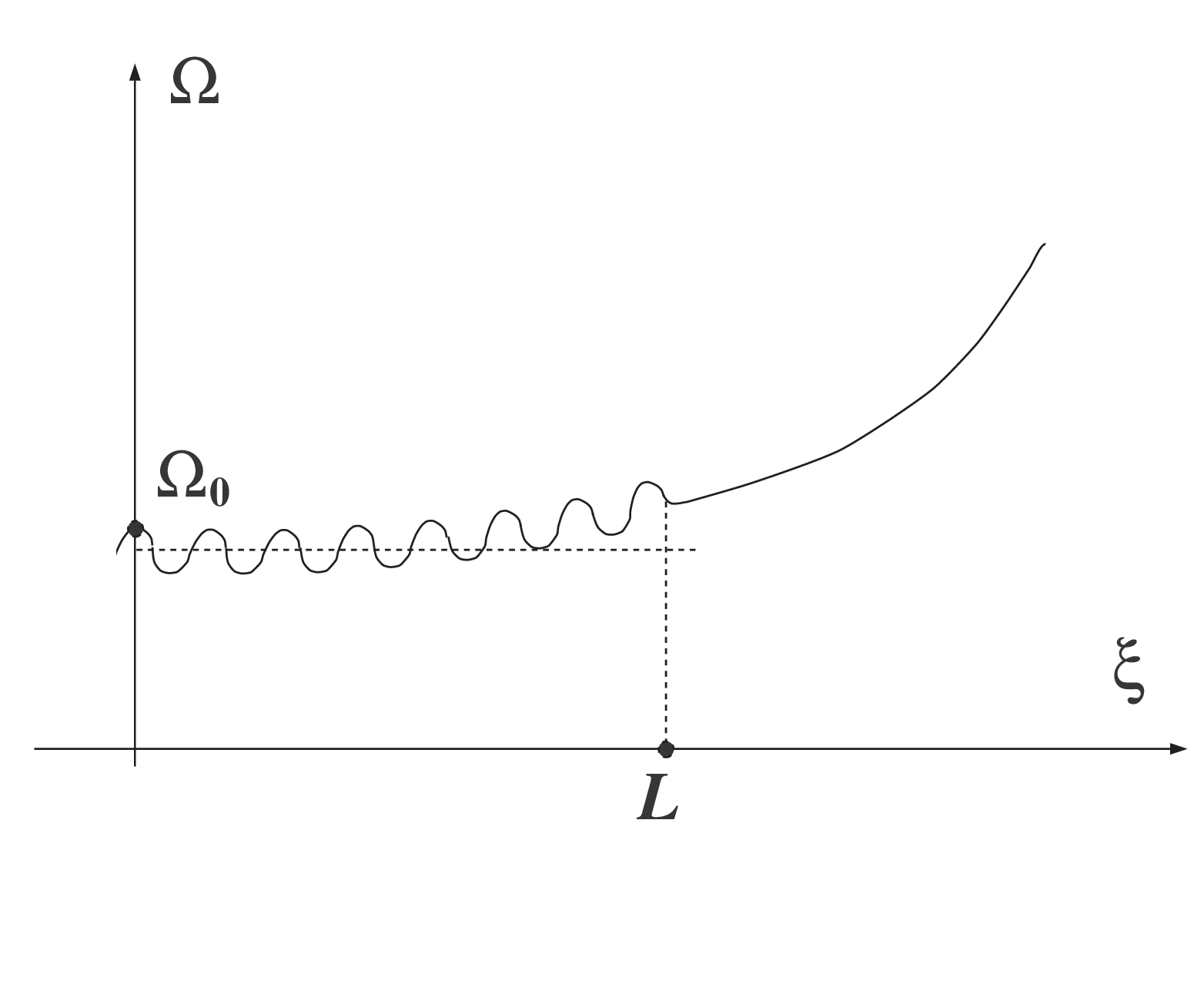}{A conformal factor $\Omega$ satisfying
requirements (i) --- (iii).} 

\section*{Acknowledgments}

I am grateful to Prof.~Grib for stimulating my studies in this field
and to Dr.~Zapatrin for useful discussion. 
\section*{References}
Anchordoqui A., Romero, G. E., Torres, D. F., and Andruchow,
\emph{I., Mod.\ Phys.\ Lett.} {\bf 14}, 791 (1999)\par\noindent 
Birrell, N.~D., and Davies, P.~C.~W., \emph{Quantum fields in curved
spacetime}, Cambridge, Cambridge University Press, 1982.\par\noindent
Epstein, H., Glaser, V., and Jaffe, A., \emph{Nuovo Cimento} {\bf
36}, 1016 (1965). 
Flamm L., \emph{Physikalische Zeitschrift} {\bf 17}, 448
(1916)\par\noindent 
Flanagan, E.~E., and Wald, R.~M., \emph{Phys.~Rev.~D} {\bf 54}, 6233
(1996).\par\noindent 
Friedman, J.~L., Schleich, K., and Witt, D.~M., \emph{Phys.~Rev.~Lett.} 
{\bf 71}, 1486 (1993).\par\noindent
Hochberg, D., Popov, A., and Sushkov, S., \emph{Phys.~Rev.~Lett.}
{\bf 78}, 2050 (1997).\par\noindent 
Krasnikov, S., \emph{Phys.~Rev.~D} {\bf 57}, 4760 (1998).\par\noindent
Krasnikov, S., Eprint \emph{gr-qc} 9909016.\par\noindent
Khatsymovsky, V., \emph{Phys.~Lett.~B\emph} {\bf 399}, 215
(1997a).\par\noindent 
Khatsymovsky, V., in \emph{Proceedings of the II Int.\ Conference
on QFT and Gravity}, TGPU Publishing, Tomsk, 1997b.\par\noindent 
Morris, M.~S.~and Thorne, K.~S., \emph{Am.~J.~Phys.} {\bf 56}, 395
(1988).\par\noindent 
Page, D.~N., \emph{Phys.~Rev.~D} {\bf 25}, 1499 (1982).\par\noindent
Sushkov, S.~V., \emph{Phys.~Lett.~A} 164, {\bf 33} (1992).\par\noindent
Taylor, B.~E., William A.~Hiscock W.~A., and Anderson P.~R.,
\emph{Phys.~Rev.~D} {\bf 55}, 6116 (1997)\par\noindent 
Visser, M., \emph{Lorentzian wormholes --- from Einstein to Hawking},
New York, AIP Press, 1995.\par\noindent 
\end{document}